\documentclass[12pt]{article}
\usepackage{amsmath}
\numberwithin{equation}{section}
\usepackage[left=0.95in,right=0.95in,bottom=1.4in]{geometry}
\usepackage{setspace}
\usepackage [linktocpage]{hyperref}
\date{(First dated: February 20, 2013)}
\begin{document}
\title{\bf Marginal Fluctuations as Instantons \\ on M2/D2-Branes \ \\ \ }
\author{{\bf M. Naghdi \footnote{E-Mail: m.naghdi@mail.ilam.ac.ir} } \\
\textit{Department of Physics, Faculty of Basic Sciences}, \\
\textit{University of Ilam, Ilam, West of Iran.}}
 \setlength{\topmargin}{0.1in}
 \setlength{\textheight}{9.2in}
  \maketitle
  \vspace{-0.0in}
    \thispagestyle{empty}
    \begin{center}
\textbf{Abstract}
\end{center}

We introduce some (anti)M/D-branes through turning on the corresponding field strengths of the eleven- and ten-dimensional supergravity theories over $AdS_4 \times M^{7|6}$ spaces, where we use $S^7/Z_k$ and $CP^3$ for the internal spaces. Indeed, when we add M2/D2-branes on the same directions with the near horizon branes of Aharony-Bergman-Jafferis-Maldacena model, all symmetries and supersymmetries are preserved trivially. In the case, we gain a localized object just in the horizon. This normalizable bulk massless scalar mode is a singlet of $SO(8)$ and $SU(4) \times U(1)$, and agrees with a marginal boundary operator of the conformal dimension of $\Delta_+=3$. However, after performing a special conformal transformation, we see that the solution is localized in the Euclideanized $AdS_4$ space and is attributable to the included anti-M2/D2-branes, which are also necessary to ensure that there is no backreaction. The resultant theory now breaks all $\mathcal{N}=8,6$ supersymmetries to $\mathcal{N}=0$ while other symmetries are so preserved. The dual boundary operator then sets up from the skew-whiffing of the representations $\textbf{8}_s$ and $\textbf{8}_v$ for the supercharges and scalars respectively, while the fermions remain fixed in $\textbf{8}_c$ of the original theory. Besides, we also address another alternate bulk to boundary matching procedure through turning on one of the gauge fields of the full $U(N)_k \times U(N)_{-k}$ gauge group in the same lines with the similar situation faced in AdS$_5$/CFT$_4$ correspondence. The latter approach covers the difficulty already faced with of the bulk-boundary matching procedure for $k=1,2$ as well.

\newpage
\setlength{\topmargin}{-0.7in}
\pagenumbering{arabic} 
\setcounter{page}{2} 

\section{Introduction}
In some recent studies \cite{Me3}, \cite{I}, \cite{N}, \cite{I.N}, we explored more the vacua of the Aharony, Bergman, Jafferis and Maldacena (ABJM from now on) model \cite{ABJM} as, by now, the best well-conjectured model of AdS$_4$/CFT$_3$ duality or M2/D2-branes theory. The main class of the solutions, which we have been searching for, are Solitons and Instantons. The instantons, which we consider, are defined as the solutions to the classical equations of motion (EOM from now on) in the Euclidean space with nonzero finite actions. As a result, there are some saddle-points contributing to the phase integral that, in turn, cause non-perturbative corrections to the main action. The way that we use here is by turning on the fluxes through the form fields while we try to keep the original ABJM background fields and geometries \cite{ABJM} unaffected. The arisen effects (and dynamics) may be considered as coming from the original branes or from the new (anti)branes added.

We first found a membrane instanton \cite{I.N}. Indeed, by making use of some ansatzs for the 4-form field strength and EOM's of the eleven-dimensional (11d from now on) supergravity over $AdS_4 \times S^7$, when $S^7$ is considered as a $S^1$ fiberation on $CP^3$, we arrived at a localized object in the bulk of Euclideanized $AdS_4$ ($EAdS_4$) space. The corresponding "irrelevant" boundary operators of the conformal dimension of $\Delta_\mp=1,2$, with the bulk "tachyon" fields, were surveyed. Then, to adjust the bulk and boundary theories, we were forced to swap the representations $\textbf{8}_s$ and $\textbf{8}_c$ of the supercharges and spinors of the ABJM model, respectively. The resultant theory was then for anti-M2-branes. Next, by turning on a singlet fermi field next to just the $U(1)$'s parts of the full gauge group, the fitted dual boundary solution with finite action was also earned.

The second exact solution was of the $U(1)$ instantons \cite{N}. Turning on some bulk modes, induced some magmatic fluctuations on the boundary. For those massless gauge fields in the bulk of Euclidean $AdS_4$, the dual boundary operators with $\Delta_\mp=1,2$ were constructed just from the $U(1) \times U(1)$ gauge fields. That solution was indeed a D0-instanton inside the original D2-branes. The electric-magnetic duality for the case and the bulk- boundary solution adjustments were also addressed. Meanwhile, another related approach to the problem was made in \cite{I}, where an uplift of the solution to the main 11d supergravity was also performed.

Next, we arrived at another fully localized solution in $EAdS_4$ \cite{Me3}. That was a pseudoscalar in the bulk coming from an anti-D4(M5)-brane whose world-volume was warping around the five (six) directions of $CP^3$ ($S^7/Z_k$). Actually, the field strength soured by the added anti-brane was a 6(7)-form written in term of the nontrivial 1-form $\omega$ on $CP^3$. The gravitational ansatz was $SU(4) \times U(1)$-invariant and so the included massless pseudoscalar was also in the already known spectra of 10d (11d) supergravity over the associated space when, of course, $S^7/Z_k$ was considered as a $U(1)$ fiberation on $CP^3$ \cite{NilssonPope}. Meanwhile, the solution broke all supersymmetries, and that the supercharges must be taken in $\textbf{8}_c$ or $\textbf{8}_v$ in contrast to the original ABJM spectra. That, in turn, signified that the resultant theory was again for anti-D2(M2)-branes because of the new included supersymmetry breaking anti-branes. So, to adjust the boundary to the bulk, we swapped the representations $\textbf{8}_s$ and $\textbf{8}_c$ of the supercharge and spinors of the ABJM model similar to that done in \cite{I.N}. The boundary "marginal" operator with the conformal dimension of $\Delta_+=3$, associated with the normalizable bulk mode, argued to have the same structure as the terms in the $SU(4) \times U(1)$-invariant Lagrangian of the ABJM first presented in \cite{ABJM} and \cite{Klebanov}. By analyzing the behavior of the bulk mode near the boundary and making use of the bulk-boundary duality rules \cite{KlebanovWitten}, \cite{Witten2}, we determined the matching field theory solution and noticed the other related issues as well. It is mentionable that, to arrive at a clear bulk solution, we needed to turn on some scalars and fermions alongside the $U(1)$ parts of the full gauge group. There, we also discussed a little on the uplifting of the 10d ansatz to 11-dimension, and argued that the solution was not at least valid for $k=1,2$, where the R-symmetry is enhanced to $SO(8)$. Another interesting hint is that we could use the boundary gauge fields to find the replying solution to the bulk. By doing so, we could say that our instanton solution might be the best counterpart to the 10d type IIB one over $AdS_5 \times S^5$ versus 4d $\mathcal{N}=4$ $SU(N)$ Yang-Mills theory \cite{Green}, \cite{Bianchi1}. In the current note we continue the lines of the studies forward.

The instanton solution in the Euclidean $AdS_4$ here has some likenesses with that in \cite{Me3} in addition to some subtle points and its origin that is almost different. Clearly, while the original 6-form and its associated bulk equation in \cite{Me3} was not invariant under a special conformal transformation, the 4-form ansatzs here in 10d and 11d supergravities over $AdS_4 \times CP^3$ and $AdS_4 \times S^7/Z_k$ respectively are conformal invariant. Indeed, we now find some massless "scalar" solutions whose conformal transformation or their skew-whiffing go into the original bulk solution in \cite{Me3}. We argue that the basic solutions preserve all supersymmetries provided that the associated M2/D2-branes are added on the same directions as the original branes in the near horizon limit of the ABJM model. The conformably transformed (or skew-whiffed or orientation-reversed) solutions break all supersymmetries as expected except when the internal space is $S^7$. Other symmetries are simply preserved and so, the corresponding boundary operator seems to be "exactly marginal". Still, an important point to say is that to not faced with the backreactions, caused by the new (anti)branes, that special conformal transformation is indeed essential.\\
On the other hand, we know that the massless scalars are in the representation $\textbf{35}_{v} \rightarrow \textbf{15}_{0} \oplus \textbf{10}_{2} \oplus \bar{\textbf{10}}_{-2}$ of $SO(8) \rightarrow SU(4) \times U(1)$ in the original ABJM spectra. So, the uncharged scalars sit in $\textbf{15}_{0}$. Nevertheless, one may construct the needed $SU(4)$-invariant $U(1)$-neutral dimension-three operators form the associated scalars in $\textbf{8}_v = \textbf{4}_{1} \oplus \bar{\textbf{4}}_{-1}$ to adjust the bulk states. Still, the main and of course interesting solution is the special skew-whiffed one, where we have to exchange the representation $\textbf{8}_s$ with $\textbf{8}_c$ or $\textbf{8}_v$ in ABJM. The appropriate swapping here is $\textbf{8}_v \rightarrow \textbf{8}_s= \textbf{1}_{2} \oplus \textbf{1}_{-2} \oplus \textbf{6}_{0}$, while the fermions remain in $\textbf{8}_c=\textbf{4}_{-1} \oplus \bar{\textbf{4}_{1}}$ unchanged. Accordingly, the scalars sit now in $\textbf{35}_{v} \rightarrow \textbf{35}_s = \textbf{1}_{0} \oplus \bar{\textbf{1}}_{4} \oplus \textbf{1}_{-4} \oplus \bar{\textbf{6}}_{2} \oplus \textbf{6}_{-2}\oplus \textbf{20}_{0}$. So, we lead again to the anti-M2/D2-brane theory by the process. Analyzing the bulk solution near the boundary, in a similar line with that in \cite{Me3}, next to our knowledge of the general forms for the marginal operators, guide us to find some fitted boundary solutions. Indeed, we use a single scalar instead of the single fermion of \cite{I.N} and also \cite{Me3}, next to setting the fermions and gauge fields to zero, to construct a field theory solution with finite action and to match the bulk-boundary facts.

Moreover, another alternative and even more proper matching elements on the boundary could be the gauge fields. So, we will use some of the gauge fields of the full $SU(N) \times SU(N)$ gauge group to construct the respective marginal operators and solutions to match with the bulk states and solutions. Actually, the corresponding operator can be constructed from the elements of the Chern-Simon terms common in M2/D2-branes theories. For simplicity, and based on some arguments, we just consider one of the $U(N)$'s and further, concentrate on the famous $SU(2)$ part. The next steps are similar to those of the matching of the well-known bulk instanton solution in $AdS_5$ with the Yang-Mills instantons on the boundary of the corresponding 4d $SU(N)$ SYM theory \cite{Bianchi1}, \cite{Chu}, \cite{Kogan}. As we will see, this new solution matches to the both common bulk solutions here and in \cite{Me3}, and especially for the undetermined cases of $k=1,2$ there.

The rest of the paper is organized as follows. In Section 2, we deal with the gravity side of the study, where we include the needed materials of the involved supergravity theories, ansatzs for the form-fields both in ten and eleven dimensions, associated supergravity spectra, clear solutions, charges, actions and other related gravitational discussions. In Section 3, the field theory side is addressed. There, we continue with the matching of the bulk solution to the boundary through the correspondence rules and find a plain boundary solution; while in the separate Section of 4, we study an alternative boundary solution by turning on the Yang-Mills gauge fields and explore how to match that boundary solution to the bulk one. Section 5 includes summary and some other related issues that may not be addressed in other parts of the paper or may be intersecting for further studies.

\section{On Gravity Side Aspects}
\subsection{Background Geometries and Fields of M2/D2-Branes, Actions and Equations of Motion}
We first review the needed materials for M2/D2-brane supergravity theories mainly from ABJM \cite{ABJM}. One always starts from M-theory over $AdS_4 \times S^7$ with $\acute{N}$ units of the 4-form flux as:
\begin{equation}\label{eq01}
ds^2_{M}=\frac{R^2}{4} ds^2_{AdS_4}+R^2 ds^2_{S^7},
\end{equation}
\begin{equation} \label{eq02}
G^{(0)}_4 \sim \acute{N} \mathcal{E}_{AdS_4},
\end{equation}
where $R$, $\acute{N}$ and $\mathcal{E}_{AdS_4}$ are the curvature radius of 11d target-space, the initial number of the flux quanta and the unit volume-form of $AdS_4$, respectively. For $AdS_4$ metric in the Poincare upper-half plane coordinates with Euclidean signature we use
\begin{equation}\label{eq03}
ds^2_{EAdS_4}=\frac{R^2}{4 u^2} \big(du^2+ dx_i dx_i \big), \quad i=1,2,3, 
\end{equation}
where $R=R_7=2R_{AdS}=2L$. For $S^7$, when considered as a $S^1$ fiber-bundle on $CP^3$, we write
\begin{equation} \label{eq04}
ds_{S^7}^2 =ds_{CP^3}^2+(d\acute{\varphi}+\omega)^2,
\end{equation}
where $\acute{\varphi}$ is the fiber coordinate with a period of $2\pi$, and $\omega$ is a topologically nontrivial 1-form related to the K\"{a}hler form $J$ on $CP^3$ (indeed $J=d\omega$) and is dual to the Reeb killing vector of $\partial_{\acute{\varphi}}$. In the last metric, the original isometry symmetry of $S^7$ breaks down as $SO(8)\rightarrow SU(4) \times U(1)$, where $SU(4)$ is the isometry of $CP^3$. After taking the $Z_k$ orbifold of $C^4$'s, which are in turn the coordinates transverse to the M2-brane world-volumes, these 8 scalars transform as $X_I \rightarrow e^{i2\pi/k} X_I $ with $I=1,2,3,4$ and $\acute{\varphi} \rightarrow \varphi/k$, and that the M-theory is now over $AdS_4 \times S^7/Z_k$ of course. When $k$ becomes large (indeed $k \rightarrow \infty$), the M-theory circle becomes small and a better description is 10-dimensional type IIA string theory over $AdS_4 \times CP^3$ with $\acute{N}=k N$ units of the 4-form flux on the quotient space. Then, we can write
\begin{equation}\label{eq05}
   ds_{ABJM(IIA)}^2 = \tilde{R}^2 \big(ds_{AdS_4}^2+4ds_{CP^3}^2 \big), \quad \tilde{R}^2 =\frac{R^3}{4k}=\pi \sqrt{2\lambda},
\end{equation}
where $\lambda\equiv N/k$ is the 't Hooft coupling of the boundary theory, and that the last approximation is valid when $\lambda\gg 1$ and also $k^5 \gg N$. We now have $N$ units of the 4-form flux $F_4^{(0)}$ on $AdS_4$ and $k$ units of the 2-form flux $F_2^{(0)}$ on the 2-cycle of $CP^1 \subset CP^3$ as
\begin{equation}\label{eq06}
   \quad F_2^{(0)}=dA_1^{(0)}=kJ, \quad F_4^{(0)}=dA_3^{(0)}=\frac{3}{8} R^3 \mathcal{E}_4, \quad H_3=dB_2=0, \quad e^{2\phi}=\frac{R^3}{k^3},
\end{equation}
where $\mathcal{E}_4$ is the unit-volume form on $AdS_4$ and $B_2$ is the NSNS 2-form of the type II theories.

We now comment on the actions from which above backgrounds arise and our solutions have to satisfy the equations of motion as well. First, we concentrate on the 10d type IIA supergravity action and equations that we are mainly working with them; whereas the 11d supergravity discussions come in its own place briefly. The 10d type IIA supergravity in string frame always reads
\begin{equation}\label{eq07}
\begin{split}
  S_{IIA} = & \frac{1}{2 \kappa^2} \int d^{10}x \, \sqrt{g} \, e^{-2\phi} \, R+\frac{1}{2 \kappa^2} \int \biggl\lbrack e^{-2\phi}  \, \bigl(4 d\phi \wedge \ast d\phi-\frac{1}{2} H_3 \wedge \ast H_3 \bigr) \\
  & -\frac{1}{2} F_2 \wedge \ast F_2-\frac{1}{2} \widetilde{F}_4 \wedge \ast \widetilde{F}_4 -\frac{1}{2} B_2 \, \wedge F_4  \, \wedge F_4 \biggr\rbrack,
\end{split}
\end{equation}
where $\widetilde{F}_4=dA_3-A_1 \wedge H_3$ and the Hodge-star operation is with respect to the full 10d metric. Among the EOM's for the form fields, metric and dilaton equations of ABJM, the dilaton equation is satisfied trivially as it takes a constant value; meanwhile, $H_3 =0$ and the Ricci-scalar $R$ vanishes for the involved geometries. The field equations that the new solutions have to satisfy, with $H_3=0$, are as well
\begin{equation}\label{eq08a}
   dF_{p}=0,\quad d\ast F_{p}=0,
\end{equation}
\begin{equation}\label{eq08b}
   d(e^{-2\phi} \ast H_3) = -F_2\wedge \ast \widetilde{F}_4+\frac{1}{2} \widetilde{F}_4 \wedge \widetilde{F}_4,
\end{equation}
where $p=2,4$. The Einstein equations of
\begin{equation}\label{eq08c}
    R_{MN}-\frac{1}{2} g_{MN} R=-8T_{MN}^\phi+ T_{MN}^{H_3}+e^{+2\phi} T_{MN}^{F_2}+e^{+2\phi} T_{MN}^{\widetilde{F}_4},
\end{equation}
where the capital indices $M, N, ...$ are for the 10d space-time directions, are also satisfied in the ABJM background. But, we should note that as long as we are not interested in changing the original background geometries, we should adjust the added fields so that the energy-momentum tensors on the right-hand-sides (RHS) of the last equation vanish. Nevertheless, as argued in the previous studies \cite{N}, \cite{Me3}, even thought there may be some small backreactions on the original backgrounds, one could simply ignore them in a probe-brane approximation or because of the holographic renormalization discussions \cite{deHaroSolodukhinSkenderis}, \cite{Bianchi3} . However, we will see that the backreactions here are small when we add the branes in the parallel directions with those of the original ABJM branes, or vanish when we add an orientation-reversed (skew-whiffed) version of the former branes, i.e. some special anti-branes.

\subsection{Ten-Dimensional Ansatzs and Solutions}
The ansatzs we consider for the 4-form field-strength of the 10d type IIA supergravity over $AdS_4 \times M^6$, with $M^6$ as a common 6d internal manifold, are
\begin{equation}\label{eq09a}
     F_4^{(1)}=df_1 \wedge A_3^{(0)} + f_1 F_4^{(0)},
\end{equation}
\begin{equation}\label{eq09b}
     F_4^{(2)}=df_2 \wedge dx \wedge dy \wedge dz,
\end{equation}
\begin{equation}\label{eq09c}
     F_4^{(3)}=d({f_3}^{-1}) \wedge dx \wedge dy \wedge dz,
\end{equation}
where $f_1,f_2,f_3$ are some scalar functions in terms of the bulk coordinates. The identity in (\ref{eq08a}) satisfies trivially and from satisfying $d\ast F_4=0$ we obtain the following differential equations
\begin{equation}\label{eq10a}
     \frac{d^2f_1(u)}{du^2} -\frac{2}{u}\frac{df_1(u)}{du} \equiv L_1 f_1(u)=0,
\end{equation}
\begin{equation}\label{eq10b}
       \frac{d^2f_2(u)}{du^2} +\frac{4}{u}\frac{df_2(u)}{du} \equiv L_2 f_2(u)=0,
\end{equation}
\begin{equation}\label{eq10c}
      L_2 f_3(u)-\frac{2}{f_3(u)} \big(\frac{df_3(u)}{du} \big)^2 \equiv L_3 f_3(u)=0,
\end{equation}
respectively. The solutions so are
\begin{equation}\label{eq11a}
     f_1(u)=c_1+c_2 u^3,
\end{equation}
\begin{equation}\label{eq11b}
     f_2(u)=c_3+\frac{c_4}{u^3},
\end{equation}
\begin{equation}\label{eq11c}
     f_3(u)=\frac{c_5 u^3}{c_6-3c_7u^3},
\end{equation}
respectively, where $c_1, c_2, ... $ are some constants related to the object charges that we encounter more later. Now, we note that the operators $L_1, L_2, L_3$ are invariant under the following conformal transformation
\begin{equation}\label{eq12}
    x_{\acute{\mu}} \leftrightarrow \frac{x_{\acute{\mu}}}{u^2+r^2},
\end{equation}
where we use $\acute{\mu}, \acute{\nu}, ...$ for the four $AdS_4$ directions and $r=\sqrt{x_i x^i}$. The conformal transformation maps a point at infinity to another at origin meanwhile it interchanges the boundary conditions. In other words, the compact space in now $S^3 \times CP^3$ with a normal vector reversed and the sign of the 4-form fluxes changes because the $\mathcal{E}_4$ sign changes by the map. For the present case, it transforms the instantons to anti-instantons actually. We come back to this point later.\\
On the other hand, the metrics of (\ref{eq01}) and (\ref{eq05}) and also the 4-form fields are invariant under the transformation. Therefore, as $f(u)$ is a solution, $f(\frac{u}{(u^2+r^2)})$ is also a solution. So, form (\ref{eq11a}), we have
\begin{equation}\label{eq13}
     f_1^t(u)\equiv f^t(u,\vec{u};0,\vec{u}_0)=c_1 + \frac{c_2 u^3}{[(\vec{u}-\vec{u}_0)^2+u^2]^3},
\end{equation}
which is indeed the boundary to the bulk propagator. With $c_7=0$, (\ref{eq11c}) leads to a same structure while we put (\ref{eq11b}) aside for now in that there is no new thing and refer the reader to \cite{LiuTseytlin} for a similar study in the type IIB theory. This conformally transformed solution is the same as that already met in \cite{Me3}, as a skew-whiffed solution, attributable to the anti-M/D-brane instantons. This solution is also a similar type with the famous D-instanton in the 10d type IIB supergravity over $AdS_5 \times S^5$ \cite{Bianchi1}, \cite{Chu}, \cite{Kogan}.\\
In order to see what happen for our ansatzs, based on the solutions and under the transformation of (\ref{eq12}), we may write
\begin{equation}\label{eq14a}
     F_4^{(1)}=\frac{3}{8} R^3 \big(f_1(u)-\frac{u}{3}\acute{f}_1(u)\big)\ \mathcal{E}_4 \Rightarrow  F_4^{(1a,1b)}=\pm \frac{3}{8} R^3 c_1 \mathcal{E}_4,
\end{equation}
\begin{equation}\label{eq14b}
     F_4^{(2)}=-u^4 \acute{f}_2(u)\ \mathcal{E}_4 \Rightarrow  F_4^{(2a,2b)}=\pm 3 c_4 \mathcal{E}_4,
\end{equation}
\begin{equation}\label{eq14c}
     F_4^{(3)}=u^4 \frac{\acute{f}_2(u)}{{f}_2(u)^2}\ \mathcal{E}_4 \Rightarrow  F_4^{(3a,3b)}=\pm \frac{3}{c_8} \mathcal{E}_4,
\end{equation}
where $\acute{}$ on $f$'s stands for the first derivative with respect to $u$, $c_8=c_5/c_6$ and the upper sign $+$ is for the original solutions while the lower sign $-$ is for the conformal transformed solutions. Note is required here that for
\begin{equation}\label{eq15a}
 c_1=1, \quad c_4=R^3/8,\quad c_8=8/R^3,
\end{equation}
the above solutions match with the original $F_4^{(0)}$ in (\ref{eq06}), and by the conformal transformation they match with its skew-whiffed version exactly.\\
So far, these included branes or anti-branes back-react on the original geometry although a negligible amount. However, we now see that for some other values of the constants they don't back-react never. We remember that on the RHS of the metric equation of (\ref{eq08c}), there are the energy-momentum tensors. In the case, we should try to cancel the competitor terms because of the new included fields in the related tensor
\begin{equation}\label{eq16}
 T_{MN}^{F_4}=\frac{1}{2.4!} \biggl[4 F_{MPQR} F_N^{PQR}-\frac{1}{2} g_{MN} F_{PQRS} F^{PQRS} \biggl],
\end{equation}
for which we just look in the $AdS_4$ components because, for the new included fields, the internal components vanish trivially. So, to nullify the backreactions, we have to set
\begin{equation}\label{eq15b}
 c_1=-2, \quad c_4=-R^3/4,\quad c_8=-4/R^3.
\end{equation}
Therefore, the requiring of no backreaction because of the added branes on the original background, leads us to a special skew-whiffing. This, in turn, means that we are indeed including some supersymmetry breaking anti-D2-branes so that the resultant theory is also for anti-D2-branes. One may also note that for the conformal transformed solutions don't back-react on the main geometry, we should take the opposite sign of the constants in (\ref{eq15b}). For both cases and all 4-forms, vanishing the energy-momentum tensors so implies that the solution we have reads
\begin{equation}\label{eq17}
 F_4=- \frac{3}{4} R^3 \mathcal{E}_4.
\end{equation}

All together, one may note that we indeed have included some electric D2-branes and anti D2-branes in the same directions with the original near horizon branes in the ABJM model. Nevertheless, the dynamics may be as some fluctuations on the main branes and not especially as from the new added branes. Anyhow, when we add D2-branes with the solutions in (\ref{eq11a}), (\ref{eq11c}) and the positive values for the constants, we have some backreactions although small; while by embedding the special anti-branes with the solution in (\ref{eq13}) and just with the special negative values of the constants in (\ref{eq15b}), we can avoid the backreactions. We should also note that for the special values of the constants in (\ref{eq15a}), the solution has the same structure as that in ABJM; while with the negative values, there is an exactly skew-whiffed version of (\ref{eq06}). Both the latter cases involve backreactions. Therefore, we mention again that to abstain from the backreactions, one may consider the interactions between the main branes of the ABJM with the special anti-branes added or between the main anti-branes with the special branes added.

Therefore, we can have two bulk theories. One that preserves all supersymmetries and one that breaks all supersymmetries. The former is when the same branes with those in the near horizon limit of ABJM are included, while the latter is when the anti-branes or some skew-whiffing of the original branes are included. In both cases, the real scalar, $f$ here, sits in $\textbf{1}_{0}$ as a singlet of $SU(4) \times U(1)$. But, we should note when one looks at the already known spectra of the involved supergravity theories on the associated spaces \cite{ABJM}, \cite{NilssonPope}, one sees that the massless scalars sit in $\textbf{35}_{v} \rightarrow \textbf{15}_{0} \oplus \textbf{10}_{2} \oplus \bar{\textbf{10}}_{-2}$ of $SO(8) \rightarrow SU(4) \times U(1)$ for the original supersymmetric $\mathcal{N}=6$ theory and also for the skew-whiffing $\textbf{8}_c \leftrightarrow \textbf{8}_s$ of the ABJM with $\mathcal{N}=0$, which we recently considered in \cite{I.N}, \cite{Me3} as well. In other words, the scalars, fermions and gravitons are originally in the representations $\textbf{8}_v$, $\textbf{8}_c$ and $\textbf{8}_s$ of $SO(8)$ in ABJM, respectively. For now and having a $SU(4)$-singlet scalar in the bulk, we should swap the representations $\textbf{8}_s$ and $\textbf{8}_v$ for the supercharges and scalars in ABJM while the fermions remain in $\textbf{8}_c$ unchanged. So, the new skew-whiffed representation and the original ones are related as
\begin{equation}\label{eq18}
  \begin{split}
  & \begin{array}{ccc}
  \left\{ \begin{split}
  \textbf{8}_v \rightarrow & \textbf{8}_s = \textbf{1}_{2} \oplus \textbf{1}_{-2} \oplus \textbf{6}_{0}, \\
  \textbf{8}_s \rightarrow & \textbf{8}_v = \textbf{4}_{1} \oplus \bar{\textbf{4}}_{-1}, \\
  & \textbf{8}_c = \textbf{4}_{-1} \oplus \bar{\textbf{4}}_{1},
  \end{split} \right. & \Rightarrow & \left\{ \begin{split}
  \textbf{35}_{v} \rightarrow & \textbf{35}_{s}=\textbf{1}_{0} \oplus \bar{\textbf{1}}_{4} \oplus \textbf{1}_{-4} \oplus \bar{\textbf{6}}_{2} \oplus \textbf{6}_{-2}\oplus \bar{\textbf{20}}_{0}, \\
  \textbf{35}_s \rightarrow & \textbf{35}_v =\textbf{10}_{2} \oplus \bar{\textbf{10}}_{-2} \oplus \textbf{15}_{0}, \\
  & \textbf{35}_c =\textbf{10}_{-2} \oplus \bar{\textbf{10}}_{2} \oplus \textbf{15}_{0},
  \end{split} \right.
  \end{array} \\
  & \ \ \ \ \ \ \ \ \ \ \ \ \ \ \ \ \ \ \ \ \ \ \ \ \ \ \ \ \ \ \ \textbf{28}_v \rightarrow \textbf{1}_{0} \oplus \bar{\textbf{6}}_{2} \oplus \textbf{6}_{-2} \oplus \textbf{15}_{0},
 \end{split}
\end{equation}
with a note that the gauge bosons sit in a same representation for all three gravitinos. Now, we have an uncharged $SU(4)$-singlet scalar ($\textbf{1}_{0}$) with the skew-whiffing of $\textbf{8}_s \leftrightarrow \textbf{8}_v$. It is mentionable that we indeed have a real scalar in the bulk because of its origin from a 3-form completely in the external space of $AdS_4$.

Here is proper to discuss on the (anti)brane charges. According to the standard formula
\begin{equation}\label{eq19a}
 Q^{D2}_e=\frac{1}{\sqrt2 \kappa^2} \int \ast F_4, \qquad Q^{D2}_m=\frac{1}{\sqrt2 \kappa^2} \int F_4,
\end{equation}
where $\kappa^2=\frac{1}{2}(2\pi)^7$, and making use of the following relations
\begin{equation}\label{eq19b}
    \ast \mathcal{E}_4= \frac{R^3}{3 k} \ J^3, \quad \ast J^3= \frac{k}{128 R^3} \ \mathcal{E}_4, \quad Vol(CP^3)=\frac{\pi^3 R^9}{6k^3},
\end{equation}
based on the solutions of (\ref{eq11a}), (\ref{eq11c}) or (\ref{eq13}), one simply obtains
\begin{equation}\label{eq19c}
 Q^{D2}_e=\frac{4}{\sqrt{2} (2\pi)^9} \frac{C}{\lambda}, \quad Q^{D2}_m=\frac{2}{\sqrt{2} (2\pi)^7} C,
\end{equation}
where the full volume of $CP^3$ and $AdS_4$ are factored out for the electric and magnetic charges, respectively; and that $C$ can be used for the overall coefficients in (\ref{eq14a}), (\ref{eq14b}) and (\ref{eq14c}). We see that the electric charge is tiny as $\lambda \gg 1$, and that in general both charges are small compared with the  background one. That is because if we use for instance the constants in (\ref{eq15b}), the resultant charge is a very small fraction of $N$ (almost zero) justifying more that the backreaction could be ignored definitely. Remember that the plus sign is for branes and the minus sign is for anti-branes here.

The respective contribution from the fifth sentence of the action (\ref{eq07}), based on the solutions, like charges, becomes
\begin{equation}\label{eq20}
 S_{inst.}^{D2}=-\frac{2}{(2\pi)^{11}} \frac{C^2}{\lambda^2},
\end{equation}
where the full 10d volume of Vol($AdS_4 \times CP^3$) is factored out as a common factor. We again see that the corrections induced by the new included (anti)branes are small really.

\subsection{On Ansatzs and Solutions in Eleven Dimensions}
The statements in the last subsections on D2-branes are also valid in the case for M2-branes, generally. More clearly, the ansatzs and solutions are valid in 11d supergravity over $AdS_4 \times S^7/Z_k$ (with $S^7/Z_k=CP^3 \times {S^1}/{Z_k}$) and satisfy the identity and equation
\begin{equation}\label{eq21}
   dF_4=0, \qquad d*_{11} F_4 + \frac{1}{2} F_4 \wedge F_4=0,
 \end{equation}
here with a note that the unit-volume form, in the 7d internal space, is
\begin{equation}\label{eq22}
   \mathcal{E}_7=\frac{1}{8.3!} J^3 \wedge e_7, \quad e^7_{{S^1}/{Z_k}}={{\frac{1}{k}}}\left(d\varphi +k\omega \right)\equiv e_7,
 \end{equation}
which next to (\ref{eq19b}) are useful in evaluating the actions and charges of the added (anti)M-branes. The issue of the backreaction is also exactly same as the former (anti)D-brane case meanwhile we don't have dilaton or any other fields here. Another point to say is that the ansatzs for $F_4$ are invariant under any internal isometric symmetry and especially $SO(8)$, which is the special case with $k=1,2$ here. \\
An interesting issue to address is the uplifting of the solution to eleven dimensions with its margins and interpretations. But, before that, one may note that the original solutions we have are pointlike in the external space similar to those in \cite{Chu}, \cite{Kogan}, and not necessarily localized in the full 10- or 11-dimensional space as that in \cite{Bianchi1}. Indeed, the solution of (\ref{eq15b}), and also that in \cite{Me3}, may be considered as smeared on $CP^3$ or $S^7/Z_k$, while the original solutions of (\ref{eq09a}), (\ref{eq09b}) and (\ref{eq09c}) are originally smeared not only on the six or seven internal directions but also on the three directions of the bulk. Namely, they are also smeared in the D2- or M2-branes world-volumes and are localized just in $u$.\\
To uplift the solution, we first note that the energy-momentum tensors, because of the new included branes, vanish exactly when the constants are those in (\ref{eq15b}) that in turn means adding some special anti-M-branes. This, in turn, means that in absence of the backreaction, the Kaluza-Klein truncation is consistent here- look, e.g., at \cite{Gauntlett} for further studies. The latter point out that the lower dimensional fields don't serve as sources for the upper dimensional ones. So, we try to construct a full 11d solution by starting from the 4d Laplace equation.\\
A solution localized in the full 11d space, with all eleven indices for the Laplace equation, can be set with an ansatz as $f(x_i,y_m)=G(u)F(\vec{u},\vec{y})$ with $\vec{y}$ considered as the eight coordinates transverse to the M2-branes world-volume and $u=\sqrt{y^m y^m}$, $m=1,\ldots,8$. In the same lines with \cite{ParkSin}, one can easily show that
\begin{equation}\label{eq23}
 \begin{split}
   & F(\vec{u},\vec{y})=c_{10}+\frac{c_{11}}{[(\vec{y}-\vec{y}_0)^2+(\vec{u}-\vec{u}_0)^2]^3}, \quad G(u)=c_{12} u^3, \\
   \Rightarrow f(u,\vec{u},\vec{y})=& c_{13}+\frac{c_{14}\ u_0^3\ u^3}{[(\vec{y}-\vec{y}_0)^2+(\vec{u}-\vec{u}_0)^2]^3}\approx \frac{u_0^3 u^3}{|\vec{X}-\vec{X}_0|^6}, \quad \vec{X}=(x^i,y^m),
 \end{split}
\end{equation}
where the eleven bosonic collective coordinates of $\vec{X}_0\equiv(x_0^i,y_0^m )$ represent the M-instanton position in the full 11d Euclidean space. This instanton solution may be a wormhole connecting the asymptotic $AdS_4 \times S^7/Z_k$ space and a flat space at the instanton location \cite{BergshoeffBehrndt}, \cite{ParkSin}. It is also notable that the solution in (\ref{eq11a}) can be considered as the $u_0 \rightarrow \infty$ limit of (\ref{eq23}) and so, (\ref{eq11a}) suits to a large instanton. On the other hand, the solution in (\ref{eq13}) is the $u_0 \rightarrow 0$ limit of this 11d localized solution and so it suits to a small instanton. We return to this later.

\section{On Field Theory Side Aspects}
\subsection{M2/D2-Branes Standard Lagrangian}
We mention the clear $SU(4)$-invariant Lagrangian of the ABJM model \cite{ABJM}, which is at hand here. For M2-branes of 11d supergravity, the near horizon geometry is $AdS_4 \times S^7$. Because of ABJM, when N stacks of these M2-branes probe a $C^4/Z_k$ singularity, the world-volume theory of the branes is a $\mathcal{N}=6$ conformal Chern-Simon matter field theory with a quiver gauge group of $U(N)_k \times U(N)_{-k}$. The matter fields transform in the bifundamental representations of the gauge group with the Chern levels of ($k, -k$). In a special limit, when the 't Hooft boundary effective coupling is $\lambda \equiv N/k \gg 1$ and $k^5 \gg N$, a better description for the gravity theory is type IIA supergravity over $AdS_4 \times CP^3$. In the procedure, $S^7$ is always considered as a U(1) fibration on $CP^3$. On the field theory side, this $U(1)$ symmetry matches to a $U(1)_b$ with $b$ for baryonic symmetry.\\
The $SU(4)_R \times U(1)_b$-invariant standard action for M2/D2-brane theories now reads
\begin{equation}\label{eq24}
 \begin{split}
   S_{ABJM}  =\int d^3x \,  \bigg\{\frac{k}{4\pi}\, \varepsilon^{\mu\nu\lambda} & \, tr \bigg(A_\mu A_\nu A_\lambda+\frac{2i}{3} A_\mu A_\nu A_\lambda-\hat{A}_\mu \hat{A}_\nu \hat{A}_\lambda-\frac{2i}{3} \hat{A}_\mu \hat{A}_\nu \hat{A}_\lambda \bigg)\\
   & -tr\big(D_\mu Y_A^\dagger D^\mu Y^A \big)-tr\big(\psi^{A\dagger} i  \gamma^\mu D_\mu \psi_A \big)-V_{bos}-V_{ferm} \bigg\},
 \end{split}
\end{equation}
where
\begin{equation}\label{eq24a}
   \begin{split}
   V_{ferm} =-\frac{2\pi i}{k} tr\big(Y_A^\dagger & Y^A \psi^{B\dagger}\psi_B - Y^A Y_A^\dagger \psi_B \psi^{B\dagger} + 2 Y^A Y_B^\dagger \psi_A \psi^{B\dagger} -2 Y_A^\dagger Y^B \psi^{A\dagger}\psi_B \\
   & +\varepsilon^{ABCD} Y_A^\dagger \psi_B Y_C^\dagger \psi_D- \varepsilon_{ABCD} Y^A \psi^{B\dagger} Y^C \psi^{D\dagger}\big),
   \end{split}
 \end{equation}
\begin{equation}\label{eq24b}
   \begin{split}
   V_{bos} =-\frac{4\pi^2}{3k^2} & tr\big(Y^A Y_A^\dagger Y^B Y_B^\dagger Y^C Y_C^\dagger + Y_A^\dagger Y^A Y_B^\dagger Y^B Y_C^\dagger Y^C + 4 Y^A Y_B^\dagger Y^C Y_A^\dagger Y^B Y_C^\dagger \\
    & - 6 Y^A Y_B^\dagger Y^B Y_A^\dagger Y^C Y_C^\dagger\big),
  \end{split}
\end{equation}
are the Bose-Fermi interaction term and the scalar bosonic potential, respectively; Note that $\mu, \nu,...$ here stand for the 3d Minkowski indices. The matter fields are four complex scalars of $Y^A$ ($A=1,2,3,4$) and four three-dimensional spinors of $\psi_A$, which transform in the bifundamental representation ($\textbf{4}_1, \bar{\textbf{4}}_{-1}$) of $SU(4)_R \times U(1)_b$. The gauge fields $A_\mu$ and $\hat{A}_\mu$ couple to the matter fields $\Phi$ ($Y^A$ or $\psi_{A}$) by the covariant derivatives
\begin{equation}\label{eq24c}
   \begin{split}
   & D_\mu \Phi =\partial_\mu \Phi+i A_\mu \Phi - i \Phi \hat{A}_\mu, \\
   & F_{\mu\nu}=\partial_\mu A_\nu-\partial_\nu A_\mu+i \big[A_\mu, A_\nu \big],
   \end{split}
 \end{equation}
where the field strength for $A$ is also given. The conventions for the metric, Clifford algebra and real gamma matrices in original Minkowski signature read
\begin{equation}\label{eq24d}
   \begin{split}
    \eta_{\mu\nu}=diag(-1,1,1), \ \ \{\gamma_\mu, \gamma_\nu \}=-2 \eta_{\mu\nu}, \ \ \gamma^\mu=(i\sigma_2,\sigma_1,\sigma_3), \ \ \varepsilon_{012}=-1,
   \end{split}
 \end{equation}
where $\sigma_{1,2,3}$ are the usual Pauli matrices. It is mentionable that the traces are taken on the $N \times N$ matrices of the gauge group keeping the gauge invariant quantities; and that the normalization for the $U(N)$ generators of $t^a$ is set as tr$(t^a t^b)=\frac{1}{2} \delta^{ab}$. When we discuss the explicit boundary solutions, we will do the continuation to the Euclidean signature.\\
This Lagrangian was first studied in \cite{ABJM}, \cite{Klebanov}, while the $\mathcal{N}=1, 2$ superfield formalism of the theory was presented in \cite{Gomiz}, \cite{Klebanov} next to some other related aspects surveyed also in \cite{Terashima}, \cite{HanakiLin}. In \cite{Raamsdonk}, it was shown that for the special case of $N=2$ M2-branes, the Bagger-Lambert-Gustavsson (BLG) theory \cite{BLG} is a special case of the ABJM theory with the gauge group of $SU(2) \times SU(2)$. Indeed, for the special cases of $k=1,2$, the $SU(4)$ R-symmetry of the ABJM is enhanced to $SO(8)$ and therefore $\mathcal{N}=8$ owing to the "monopole operators" \cite{GustavssonRey}.

\subsection{Matching the Bulk Solutions to the Boundary}
A scalar field in the bulk of the Euclidean $AdS_4$, when approaching the boundary at $u=0$ of the Poincare upper-half plane coordinates (\ref{eq03}), behaves like \cite{KlebanovWitten}
\begin{equation}\label{eq25}
    f(u,\vec{u}) \approx \alpha(\vec{u})\ u^{\Delta_-} + \beta(\vec{u})\ u^{\Delta_+},
\end{equation}
where $\Delta_\mp$ are the roots of $(m L)^2=\Delta(\Delta-3)$. For a massless scalar, $\Delta_\mp=0,3$ from which we use $\Delta_+=3$ that corresponds to the normalizable mode in the bulk. $\alpha$ and $\beta$ play the roles as the "source" and "vacuum expectation value" of the marginal operator of the conformal dimension of $\Delta_+=3$, respectively and vice versa for $\Delta_-$ \cite{KlebanovWitten}, \cite{Balasubramanian}.\\
Such a scalar can be quantized either with Dirichlet boundary condition $\delta\alpha=0$ (which can be used for any $m^2$) or with Neumann or mixed boundary condition $\delta\beta=0$ (which can be used when the scalar masses are in the range of $-9/4<m^2 L^2<-5/4$, ensuring stability too). For the Dirichlet boundary condition, the stability needs the scalar mass obeys the Breitenlohner-Freedman (BF) bound $m^2 L^2 \geq -9/4$ \cite{BreitenlohnerFreedman}. These boundary conditions preserve the asymptotic symmetry groups, lead to finite energies and correspond to two boundary CFT's. The "usual" CFT is that for which a source $\alpha$ couples to an operator of the conformal dimension of $\Delta_+$.\\
Anyway, we now note that for the solution in (\ref{eq11a}), both $\alpha$ and $\beta$ are constants, and for which we propose an operator of the conformal dimension of 3. Meanwhile, for the solution of (\ref{eq13}), the procedure here is the same as that outlined in \cite{Me3}. Indeed, by comparing (\ref{eq13}) with (\ref{eq25}), we can write
\begin{equation}\label{eq26a}
    \alpha(\vec{u})=f_0(\vec{u}), \qquad \beta(\vec{u})=\frac{c_2}{|\vec{u}-\vec{u}_0 |^6}.
\end{equation}
For a localized object on the boundary, the source of $f_0(\vec{u}_0)$ is indeed a delta function $\delta^3(\vec{u}-\vec{u}_0)$ and we have
\begin{equation}\label{eq26b}
   \frac{1}{3} \langle \mathcal{O}_3(\vec{u})\rangle_\alpha=-\frac{\delta W[\alpha]}{\delta\alpha}= \beta(\vec{u}),
\end{equation}
in which $W=-S_{on-shell}$ are, from left, the field theory "generating functional" and the bulk "on-shell action" that we evaluate bellow. Then, in the language of \cite{Witten2}, because of turning on the normalizable bulk scalar mode, we should deform the action as $S \rightarrow S + W$, where
\begin{equation}\label{eq26c}
     W= -\frac{1}{3} \int d^3 \vec{u} \ \alpha(\vec{u})\ \mathcal{O}_3(\vec{u}),
\end{equation}
with a note that $\alpha=c_1$ here, and that the plain forms for the operators come in the next subsection and section.

\subsection{Boundary Solutions and Correspondence}
First we remember that our ansatzs for $F_4$ are $SU(4)$ and $SO(8)$ invariant and actually singlet in addition that they don't carry any $U(1)$ charge. So, the dual operator $\mathcal{O}_3$ have to have the same property. Nevertheless, we already know that the normalizable mode may be considered as a different sate in the original theory and not necessarily as a deformation of that \cite{Balasubramanian}. This fact suggests that the dual operators have the same structures as the main Lagrangian (\ref{eq24}) terms. This statement is confirmed in some previous studies on the spectra and Bogomol'nyi-Prasad-Sommerfeld (BPS) operators to which the bulk modes agree \cite{AharonyOzYin}, \cite{Minwalla}, \cite{Halyo1},\cite{HokerPioline}. The common proposed marginal operator, dual to the bulk scalars, can be
\begin{equation}\label{eq27a}
     \acute{\mathcal{O}}_3= tr\big(X^{[I} X_J^\dagger X^{K]} X_{[K}^\dagger X^J X_{I]}^\dagger\big),
\end{equation}
where the suitable trace subtraction is mentionable. This operator has, of course, the same structure with the scalar sextet potential of the BLG and ABJM model.Indeed, we should note that if one uses $Y^A=X^A + i X^{A+4}$, the ABJM scalar potential of (\ref{eq24b}) coincides with the BLG one, which is $-\frac{32\pi^2}{3k^2}\acute{\mathcal{O}}_3$ \cite{Raamsdonk}, \cite{Klebanov}.  \\
On the other hand, we recall that we have two types of solutions. One may be attributed to adding some M2/D2-branes in the same directions with the original ones in the near horizon of the ABJM and thus preserving all symmetries and supersymmetries; While the other one may be attributed to adding some anti-M2/D2-branes with a flipped direction (orientation-reversed) with respect to the original ones because of the conformal transformation of (\ref{eq12}), which preserves all symmetries but breaking all supersymmetries. Now, we try to make solutions for both cases.\\
By the way, we know that the massless scalars in the original ABJM theory sit in the representation $\textbf{35}_{v} \rightarrow \textbf{15}_{0} \oplus \textbf{10}_{2} \oplus \bar{\textbf{10}}_{-2}$ of $SO(8) \rightarrow SU(4) \times U(1)$ with $X^I \rightarrow (Y^A, Y_A^\dagger)$. This marks that the $U(1)$-neutral ones are in $\textbf{15}_{0}$ while our bulk scalars are singlet and sit in $\textbf{1}_{0}$. But, no problem! Because the representation for $Y^A Y_A^\dagger$ is $\textbf{4}_{1} \otimes \bar{\textbf{4}}_{-1}=\textbf{15}_{0}+\textbf{1}_{0}$ and so, form (\ref{eq27a}), one can easily see that there is a $SU(4)_R \times U(1)_b$-singlet. Therefore, with a more intimately operator, of the conformal dimension of $\Delta_+=3$, composed of the ABJM scalars like
\begin{equation}\label{eq27b}
     \mathcal{O}_3= tr\big(Y^A Y_A^\dagger Y^B Y_B^\dagger Y^C Y_C^\dagger\big),
\end{equation}
to agree with a bulk solution like (\ref{eq11a}), with respect to (\ref{eq26b}) and (\ref{eq26c}), one may simply set
\begin{equation}\label{eq28}
     Y^A=Y_A^\dagger= c_{15} \emph{\textbf{I}}_{N \times N},
\end{equation}
where $\emph{\textbf{I}}_{N \times N}$ is the unitary matrix. We see that there is no critical point for the case. But, for the conformally transformed or the skew-whiffed solution in (\ref{eq13}), which corresponds to the anti-M2/D2-brane theories, the situation is intersecting. \\
As discussed in (\ref{eq18}), for the skew-whiffed bulk solution, we should swap the representations $\textbf{s}$ and $\textbf{v}$ for the supercharges and scalars of the ABJM model. The scalars now sit in $\textbf{8}_v \rightarrow \textbf{8}_s = \textbf{1}_{2} \oplus \textbf{1}_{-2} \oplus \textbf{6}_{0}$ according to which we break up the scalar fields as $X^I \equiv (y^n, y^7, y^8) \approx (y^n, y, \bar{y})$ with $n=1,...,6$, $y=y^7+iy^8$, and for simplicity also with $\bar{y}=y^\dagger$. The situation is similar with \cite{I.N}, \cite{Me3}, where we had a single fermion instead of a scalar here. By making use of just this single scalar and setting other six scalars to zero, the marginal operator, from (\ref{eq27b}), reads
\begin{equation}\label{eq29}
     \mathcal{O}_3= tr (y\bar{y})^3,
\end{equation}
while there are still other ways to construct such an operator from the scalars. It is simple to see that with $y^n=0$, the scalar potential from the action of (\ref{eq24}) vanishes. Afterwards, by setting the gauge fields and also fermions to zero and with respect to (\ref{eq26c}), the remaining part of the Lagrangian reads \footnote{It is notable that one could simply set all fields in the action of (\ref{eq24}) to zero except one real scalar field. One then gets the wished solution trivially without dealing with any deformation as the marginal normalizable bulk mode suggests that the boundary solution may be from the main action and not necessarily from any deformation of the main action.}
\begin{equation}\label{eq30}
     \mathcal{L}_{inst.}=-tr(\partial_i y \partial_i \bar{y}) - \frac{c_1}{3} tr (y\bar{y})^3.
\end{equation}
Next, by taking $y=i h(r)\textbf{1}_{1 \times 1}$, we arrive at
\begin{equation}\label{eq30a}
    h(r)= \sqrt[1/2]{\frac{3}{c_1}} \bigg(\frac{\grave{c}}{\grave{c}^2+r^2}\bigg)^{1/2},
\end{equation}
with $\grave{c}$ another constant. Then, based on the current solution with $c_1=1$, the finite part of the action becomes
\begin{equation}\label{eq30b}
   S_{inst.S}= \frac{\sqrt{3}}{2} \pi^2.
\end{equation}
Now, as a main test of the duality correctness, we see that the one-point function of the involved operator $\mathcal{O}_3(\vec{u})$, based on the solution of (\ref{eq30a}), is the same as $\beta(\vec{u})$ in (\ref{eq26a}) up to some constants, which the constants in turn adjust together.

To sum up the subsection, we remind the reader that our bulk solutions with $\mathcal{N}=6,0$ supersymmetries have some equivalents in the ABJM $SU(4)_R \times U(1)_b$-invariant Lagrangian especially for $k \geq 3$ when we use the Hopf-fibred $S^7/Z_k=CP^3 \times S^1/Z_k$ in the correspondence with the already known spectra \cite{Halyo2}. For $k=1,2$, our bulk solution is tantamount to the boundary solution with $SO(8)$ R-symmetry and so $\mathcal{N}=8$ supersymmetry. A noticeable point is that as long as the ABJM scalar potential, which we use to match the solutions, is $SO(8)$-invariant \cite{GustavssonRey}, \cite{Raamsdonk}; we can implicitly say that the solutions are valid for all $k$'s. Of course, we should note that, in general, the $SO(8)$ symmetry combines the standard operators like $tr(Y^AY_A^\dagger)^\ell$ with the monopole-operators to enhance the symmetry \cite{ABJM}.

Nonetheless, we may still go another way to match the bulk to the boundary solutions. That is using the other common terms in the BLG and ABJM Lagrangian's. So, we use the Chern-Simon terms in the next section. We will see that the handling of the gauge fields is meaningful both in finding the boundary solutions for $k=1,2$ cases, which were undetermined in \cite{Me3}, and also in following the lines already used to find the D-instanton in 10d type IIB theory over $AdS_5 \times S^5$ versus 4d $SU(N)$ $\mathcal{N}=4$ Yang-Mills field theory.

\section{The Bulk Solution from Chern-Simon Action}
We now consider just the universal Chern-Simon terms and gauge fields of the action (\ref{eq24}) for M2/D2-branes. For convenience and sake of the solution matching, we put away one group of the full $U(N)_k \times U(N)_{-k}$, which in turn means using one of the gauge fields, say $A_i$. Still, an even more logical reason, for doing so, may come from a "novel Higgs mechanism", where the quiver gauge group of $SU(N) \times SU(N)$ breaks down to $SU(N)$ under certain conditions; look at \cite{ChuNastaseNilssonPapageorgakis} and references therein for further related studies. Another reason could be the breaking of the parity by the supersymmetry breaking solution of \cite{Me3}, which on related situations some typical studies are also done in \cite{FujitaLiRyuTakayanagi}. Anyhow, we simply take $\hat{A}_i=-A_i$. So, the remaining Lagrangian with setting the scalars and fermions to zero reads
\begin{equation}\label{eq31}
   \mathcal{L}_{CS}=\frac{k}{4\pi} tr \bigg(A \wedge dA +\frac{2i}{3} A \wedge A \wedge A \bigg),\quad F=dA + i A \wedge A.
\end{equation}
On the other hand, we know that to have a finite action by a configuration, the Lagrangian density should be nonzero just in a localized region of the space and vanish at the boundary of the Euclidean space. It means that for the condition of $F_{ij}=0$ here, $A_i$ should behave as a pure gauge at infinity or origin. However, it is well-known that all gauge fields with vanishing field strengths at infinity can be classified by an integer number $K$ called "instanton number" or "Pontryagin index" \cite{Belavin}
\begin{equation}\label{eq32a}
   K=\frac{1}{16\pi^2} \int_{R^4} d^4x\ tr(F_{\acute{\mu}\acute{\nu}}\ast_4 F^{\acute{\mu}\acute{\nu}}), \quad \ast_4 F_{\acute{\mu}\acute{\nu}}=\frac{1}{2} \mathcal{E}_{\acute{\mu}\acute{\nu}\acute{\rho}\acute{\sigma}} F^{\acute{\rho}\acute{\sigma}},
\end{equation}
by noting of the conformal flatness of the Euclidean $AdS_4$. By using the Stocks-theorem, the integral on $R^4$ can be reduced to an integral on $\partial R^4 \approx S^3$. So, for our interest, we have
\begin{equation}\label{eq32b}
   K=\frac{1}{8\pi^2} \int_{R^3\approx S^3} d^3x\ \varepsilon^{ijk}\ tr \bigg(A_i \partial_j A_k+\frac{2i}{3} A_i A_j A_k\bigg),
\end{equation}
which is of the same structure that we have in the action. Now, to adjust with the bulk solution in (\ref{eq13}) and \cite{Me3}, we use the special "singular" gauge
\begin{equation}\label{eq32c}
   A_i=\frac{u^2}{r^2+u^2} g^{-1} \partial_i g, \quad g=\frac{(u{1}_2 -i x_i \sigma^i)}{(r^2+u^2)^{1/2}},
\end{equation}
where the gauge field $A_i$ behaves as a pure gauge on the boundary ($\vec{u}\rightarrow \vec{u}_0$) and $g$ is an element of $SU(N)$. We use here the simplest case $SU(2)$ of course. This simple case is even more relevant in that with $k=1,2$ and $SO(8)$ R-symmetry, the gauge group of BLG is indeed $SU(2) \times SU(2)$ \cite{Raamsdonk}, \cite{Klebanov}, which is in turn the case with two M2-branes in ABJM \cite{ABJM}. \\
Now, by evaluating the respective part of the action in (\ref{eq24}) and by hinting that it picks up a $i$ factor because of the Wick-rotation to the Euclidean space, we have
\begin{equation}\label{eq32d}
   S_{CS}=\frac{k}{6\pi} \int_{S^3}\ tr\big(A^3\big), \quad S_{inst.YM}=4\pi k,
\end{equation}
where we have taken the integral on the north pole ($u=1, \vec{u}=0$) of the unit-sphere by noting that $F=0$ for pure gauge. We see this is the same value with that got in \cite{Me3} by another way and also with that in (\ref{eq24}) nearly. Note also that the topological charge here is $K=-1$.

As an other side, we try to follow the similar lines with \cite{Kogan}, \cite{ParkSin}, where the instanton was indeed a D(-1)-brane inside the world-volume of the N background D3-branes. To do so, we note that the instanton charge we have, because of the added anti-M2/D2-branes, is
\begin{equation}\label{eq33}
   Q_{inst.}= \int_{AdS_4 \times CP^3|S^7/Z_k} \ \ast(d\ast df),
\end{equation}
where $f$ is that in (\ref{eq13}) and the Hodge-star operation just here is in 10 or 11 dimensions. By noting that the last integrand is indeed the Laplace equation, we get
\begin{equation}\label{eq33a}
  Q_{inst.}=\frac{c_2}{3Vol(\partial AdS_4)} \lim_{u\rightarrow0}  \int_{R^3} d^3x\  \frac{1}{u^3} \frac{u^3}{(u^2+r^2)^3}=\frac{c_2}{3},
\end{equation}
where the internal-space volume is factored out. \\
On the other hand, we recall that the conformal transformation of (\ref{eq12}) allows us to relate the behavior of the system at origin and at infinity, where the associated asymptotic spaces may connect by a throat. In other words, the original solution (\ref{eq11a}) for the branes is singular at infinity in contrast to its conformally transformed (\ref{eq13} for anti-branes, which is singular at $u=0$. The latter corresponds to a small instanton on the boundary that leaves the background geometry unchanged. Similar to D-instanton in type IIB over $AdS_5 \times S^5$ in string frame, the configuration is composed of two asymptotic $AdS_4 \times S^7$ spaces, one at singularity at origin and another at infinity, which are connected by a throat \cite{BergshoeffBehrndt}, \cite{ParkSin}. One interpretation may be that we here have branes at infinity and anti-branes on the boundary, which are connected by a throat from which the instanton charge flows. With this language, the instanton number of $K$ in (\ref{eq32b}) may identify with the instanton charge of $Q_{inst.}$ as
\begin{equation}\label{eq34}
  K\approx Q_{inst.}\approx \lim_{u\rightarrow0}  \int_{R^3} d^3x\  \frac{1}{u^3} \frac{u^3}{(u^2+r^2)^3}.
\end{equation}
In is mentionable that the solution of (\ref{eq13} matches to a small Yang-Mills instanton when we approach the boundary at $u=0$. In other words, three coordinates of $\vec{u}_0$ describe the instanton location, while the scalar parameter $u_0$ is for the instanton size. They are indeed four moduli space parameters of $SO(4,1)/SO(4)$, where $SO(4,1)$ is the isometry group of Euclidean $AdS_4$ and $SO(5)$ is a subgroup consists of rotations, translations and special conformal transformations keeping invariant the instanton up to some gauge transformations. We should also note that one can gain $SU(N)$ instantons by some embedding of the $SU(2)$ instantons into $SU(N)$ so that various embeddings lead to various configurations \cite{BelitskyVandorenNieuwenhuizen}.

Further, we note that because of the D-instanton presence on D3-branes, the gauge field $F$ on the D3-brane world-volume couples to the axion field of $A_0$ with $F_1=dA_0$ as $S_{WZ}\sim \int A_0 F \wedge F$ \cite{Chu}, \cite{Kogan}, \cite{Bianchi1}. What we can say here? We may roughly name the object as some M- or D-instanton; but a more probable counterpart may be $F_0=df$ as in $S_{WZ}\sim \int F_0 A \wedge A \wedge A$. Indeed, by taking the 10- or 11-dimensional Hodge-star on the latter 1-form and then taking the exterior derivative, the $AdS_4$ part of the resultant 10-form is the integrand in (\ref{eq33}). In a known language, this 0-form is named as "Rommans mass" which plays here a similar role as if it was a D(-2)-brane in type IIA like the D(-1)-brane in type IIB. Indeed, it was already shown in \cite{GaiottoTomasiello} that there are such $\mathcal{N}=0$ supersymmetry breaking solutions of ABJM.

\section{Summary and Comments}
This work is on the trail of the recent studies \cite{I.N}, \cite{N}, \cite{I}, \cite{Me3} in searching to find the vacua and mainly instantons of the M2/D2-branes theories in the framework of AdS$_4$/CFT$_3$ correspondence. We applied some 4-form field strengths of the 10- and 11-dimensional supergravities in terms of $AdS_4$ ingredients completely to stand some solutions in the bulk. Actually, when the equations of the motion applied to the ansatzs, we had a solution localized just in the horizon, which was of course associated with some special branes included on the same directions with the background ones. As the solutions were invariant under a special conformal transformation, we then gained a special skew-whiffed solution completely localized in the Euclidean $AdS_4$ avoiding the backreaction as well. The solution, also in \cite{Me3}, was associated with the branes with flipped directions (orientation-reversed) with respect to the main M2/D2-branes in the near horizon of the ABJM model \cite{ABJM} that broke all supersymmetries while preserving other symmetries of the original theory. Because the bulk normalizable mode was a $SO(8)$- and $SU(4)$-singlet massless scalar neutral under $U(1)$, the dual boundary operator of the conformal dimension of $\Delta_+=3$ should be also the same singlet. \\
On the other hand, from the known spectra of the gauged supergravities on the associated spaces $AdS_4 \times S^7/Z_k| CP^3$ \cite{NilssonPope}, one could see when the gravitinos were in $\textbf{8}_{s}$ and $\textbf{8}_{c}$, the uncharged scalars were so in $\textbf{15}_0$ of $SU(4)$; while for the gravitino of $\textbf{8}_v$, they were in both $\textbf{1}_0$ and $\textbf{20}_0$. However, this was invalid for $k=1,2$ of the quotient space of $S^7/Z_k$ \cite{Halyo2}. The form of such marginal operators were of the same type with the sentences in the main ABJM Lagrangian (\ref{eq24}) \cite{Balasubramanian}, suspected also in \cite{AharonyOzYin}, \cite{Minwalla}, \cite{Halyo1}. For the current case, they were in turn of the same structure with the bosonic potential of (\ref{eq24b}). For the proposed added branes, the operator was trivial; meanwhile for the anti-branes we should swap the representations $\textbf{8}_{s}$ and $\textbf{8}_{v}$ for the supercharges and scalars in ABJM while the fermions remained fixed in $\textbf{8}_{c}$. After that, we used a single scalar field to construct the dual boundary operator and then, by making use of the known AdS/CFT duality rules \cite{KlebanovWitten}, \cite{Witten2}, we matched the bulk solution to a clear boundary one. In addition, we used another possibility to match with the fully localized bulk object, suitable for the $k=1,2$ cases particulary. That was from the Yang-Mills fields of the common Chern-Simon terms of the standard M2/D2-branes Lagrangian \cite{Raamsdonk}, \cite{Klebanov}, \cite{ABJM}. By keeping just one of the gauge group and then a $SU(2)$ subgroup, we found a dual boundary solution equivalent to the instanton of the AdS$_4$/CFT$_3$ duality \cite{Bianchi1}, \cite{Chu}, \cite{ParkSin}, \cite{Kogan}, where the Rommans mass $F_0$ here might play a similar role as the axion $A_0$ there.

Another point to say is about the magnetic dual of these new (anti)D2- and M2-branes. Indeed, we note that the magnetic dual of the 4-forms here couple to (anti)D4-branes and (anti)M5-branes, respectively. Then, the world-volumes of such (anti)branes supposed to wrap around some of the five and six directions of $J^3$ and $J^3 \wedge e_7$, respectively. That is a similar situation already encountered in \cite{Me3}, where an anti-D4-brane supposed to wrap around some parts of $J^2 \wedge \omega$ and its 11d counterpart supposed to wrap around some parts of $J^2 \wedge \omega \wedge d\varphi$.\\
Finally, we saw that for a special combination of the constants in the solutions, the backreactions were ignored logically. Therefore, we may argue that the bulk scalars don't disturb the upper dimensional fields as it is thought that the truncation is now consistent \cite{Gauntlett}, \cite{deHaroSolodukhinSkenderis}. However, although the supersymmetry always ensures stability and that the skew-whiffed solution is stable just for $S^7$, for the similar situations, the stability is already surveyed in \cite{DuffNilssonPope} after \cite{BreitenlohnerFreedman}. In fact, it is argued in \cite{DuffNilssonPope} that the skew-whiffed Freund-Rubin type solutions are at least perturbatively stable in the large N limit, and that for $k\geq2$ the stability is guaranteed with $S^7/Z_k$. Nevertheless, it might be a probable instability originated in the $0^+$ sector. Finding out whether or not the existing solution is stable needs a clear evaluation with a note that the instanton is responsible for tunneling between the original M2/D2-branes and the orientation-reversed anti-M2/D2-branes.\\
On the other hand, because of the "tachyon condensation" as a result of the brane-antibrane pair formations, the systems may be unstable \cite{Sen} as the tachyons also present a "flip transition" generally \cite{Narayan}. Still, one should note that the marginal operators appearing in the skew-whiffed nonsupersymmetric theories may disturb the conformal fixed points and produce instabilities \cite{Berkooz}. So, in general, an outspoken analysis and evaluation of the quantum corrections produced by the solutions are required to settle certainly if the system is stable.

\end{document}